\documentclass[12pt]{iopart}
\usepackage{iopams}
\usepackage{epsfig}
\begin{document}

\title[High-frequency sources of gravitational waves]{High frequency sources of gravitational
waves}

\author{Kostas D Kokkotas \dag}

\address{\dag\ Department of  Physics, Aristotle University of Thessaloniki, 54124 Greece}

\begin{abstract}
Sources of high frequency gravitational waves are reviewed.
Gravitational collapse, rotational instabilities and oscillations
of the remnant compact objects are potentially important sources
of gravitational waves. Significant and unique information for the
various stages of the collapse, the evolution of protoneutron
stars and the details of the equations of state of such objects
can be drawn from careful study of the gravitational wave signal.
\end{abstract}



%\maketitle

\section{Introduction}

The new generation of gravitational wave (GW) detectors is already
collecting data by improving the sensitivity by at least one order
of magnitude compared to the operating resonant detectors.
Broadband GW detectors are sensitive to frequencies between 50 and
a few hundred Hz. The next generation will broaden the bandwidth
but it will still not be sensitive enough to frequencies over
500-600Hz, unless they are operated as narrow-band detectors
\cite{ThorneCutler,GEO600}. There are also suggestions for wide
band resonant detectors in the kHz band \cite{Cerdonio}.

In this short review we will discuss some of the sources that are
in the frequency band ($\gtrsim 500-600$Hz), where the
interferometers are sensitive enough only if they are
narrowbanded. Since there is a variety of GW sources with very
interesting physics associated to them, this high-frequency window
deserves special attention. If either resonant or narrow-band
interferometers achieve the needed sensitivity, there is a
plethora of unique information that can be collected.

\section{Gravitational collapse}

One of the most spectacular events in the Universe is the
supernova (SN) collapse to create a neutron star (NS) or a black
hole (BH). Core collapse is a very complicated event and a proper
study of the event demands a deeper understanding of neutrino
emission, amplification of the magnetic fields,  angular momentum
distribution, pulsar kicks, etc. There are many viable
explanations for each of the above issues but it is still not
possible to combine all of them together into a consistent theory.
Gravitational waves emanating from the very first moments of the
core collapse might shed light on all the above problems and help
us understand the details of this dramatic event. Gravitational
collapse compresses matter to nuclear densities, and is
responsible for the core bounce and the shock generation. The
event proceeds extremely fast, lasting only less than a second,
and the dense fluid undergoes motions with relativistic speeds
($v/c\sim 0.2-0.4$). Even small deviations from spherical symmetry
during this phase can generate copious amounts of GWs. However,
the size of these asymmetries is not known.
 From observations in the
electromagnetic spectrum we know that stars more massive than
$\sim 8M_\odot$ end in core collapse and that $\sim 90\%$ of them
are stars with masses $\sim 8-20M_\odot$. During the collapse most
of the material is ejected and if the progenitor star has a mass
$M\lesssim 20M_\odot$ it leaves behind a neutron star. If
$M\gtrsim 20M_\odot$ more than 10\% falls back and pushes the
proto-neutron-star (PNS) above the maximum NS mass leading to the
formation of black holes ({\em type II collapsars}). Finally, if
the progenitor star has a mass $M\gtrsim 40M_\odot$ no supernova
is launched and the star collapses to form a BH ({\em type I
collapsars}).

A significant amount of the ejected material can fall back,
subsequently spinning up and reheating the nascent NS.
Instabilities can be excited again during such a process. If a BH
formed its quasi-normal modes (QNM) can be excited for as long as
the process lasts. ``Collapsars'' accrete material during the very
first few seconds, at rates $\sim 1-2M_\odot$/sec. Later the
accretion rate is reduced by an order of magnitude but still
material is accreted for a few tenths of seconds. Typical
frequencies of the emitted GWs are at the range 1-3kHz for $\sim
3-10 M_\odot$ BHs. If the disk around the central object has a
mass $\sim 1M_\odot$ self-gravity becomes important and
gravitational instabilities (spiral arms, bars) might develop and
radiate GWs. There is also the possibility that the collapsed
material might fragment into clumps, which orbit for some cycles
like a binary system ({\em fragmentation instability}).

The supernova event rate is 1-2 per century per galaxy
\cite{Cappellaro} and about 5-40\% of them produce BHs through the
fall back material \cite{FryerKalogera}. Conservation of angular
momentum suggests that the final objects should rotate close to
the mass shedding limit, but this is still an open question, since
there is limited knowledge of the initial rotation rate of the
final compact object. Pulsar statistics suggest that the initial
periods are probably considerably shorter than $20$ms.  This
strong increase of rotation during the collapse has been observed
in many numerical simulations (see e.g. \cite{FryerHeger,
DFM2002b}).

Core collapse as a potential source of GWs has been studied for
more than three decades (some of the most recent calculations can
be found in \cite{Zwerger1997, Rampp1998, Fryer2002, DFM2002b,
Ott2003, Kotake2003}). All these numerical calculations show that
signals from Galactic supernova ($d\sim 10$kpc) are detectable
even with the initial LIGO/Virgo sensitivity at frequencies
$\lesssim$1kHz. Advanced LIGO can detect signals from distances of
1Mpc but it will be difficult with the designed broadband
sensitivity to resolve signals from the Virgo cluster
($\sim$15Mpc). The typical GW amplitude from the 2D numerical
simulations \cite{DFM2002b, Ott2003} for an observer located on
the equatorial plane of the source is
\begin{equation}
h\approx 9 \times 10^{-21}\varepsilon \left({ 10 {\rm kpc} \over
d}\right)
\end{equation}
where $\varepsilon \sim 1$ is the normalized GW amplitude. The
total energy radiated in GWs during the collapse is $\lesssim
10^{-6}-10^{-8} M_\odot c^2$. These numerical estimates are not
yet conclusive, important aspects such as 3D hydrodynamics
combined with proper spacetime evolution have been neglected. The
influence of the magnetic fields have been ignored in most
calculations. The proper treatment of these issues might not
change the above estimations by orders of magnitude but it will
provide a conclusive answer. There are also issues that need to be
understood such as the pulsar kicks (velocities even higher than
1000 km/s) which suggest that in a fraction of newly-born NSs (and
BHs) the process may be strongly asymmetric, or the polarization
of the light spectra in supernovae which also indicate significant
asymmetries \cite{Wheeler1999}. Better treatment of the
microphysics and construction of accurate progenitor models for
the angular momentum distributions are needed. All these issues
are under investigation by many groups.

%%%%%%%%%%%%%%%%%%%%%%%%%%
\section{Rotational instabilities}

Newly born neutron stars are expected to rotate rapidly, being
subject to rotation induced instabilities. These arise from
non-axisymmetric perturbations having angular dependence $e^{i
m\phi}$. Early Newtonian estimates have shown that a {\em
dynamical bar-mode ($m=2$) instability} is excited if the ratio
$\beta=T/W$ of the rotational kinetic energy $T$ to the
gravitational binding energy $W$ is larger than $\beta_{\rm dyn}=
0.27$. The instability develops on a dynamical time scale (the
time that a sound wave needs to travel across the star) which is
about one rotation period and may last from 1 to 100 rotations
depending on the degree of differential rotation in the PNS.
Another class of instabilities are those driven by dissipative
effects such as fluid viscosity or gravitational radiation. Their
growth time is much longer (many rotational periods) but they can
be excited for significantly lower rotational rates, $\beta
\gtrsim 0.14$.

\subsection{Bar-mode instability}

The bar-mode instability can be excited in a hot PNS, a few
milliseconds after the core-bounce, given a sufficiently large
$\beta$. It might also be excited a few tenths of seconds later,
when the NS cools enough due to neutrino emission and contracts
still further ($\beta \sim 1/R$). The amplitude of the emitted
gravitational waves can be estimated as $h\sim M R^2 \Omega^2/d$,
where $M$ is the mass of the body, $R$ its size, $\Omega$ the
rotational rate and $d$ the distance from Earth. This leads to an
estimation of the GW amplitude
\begin{equation}
h\approx 9\times 10^{-23} \left({\epsilon \over 0.2} \right)
\left({f\over 3 {\rm kHz}}\right)^2 \left({15 {\rm Mpc} \over
d}\right) M_{1.4} R_{10}^2.
\end{equation}
where $\epsilon$ measures the ellipticity of the bar. Note that
the GW frequency $f$ is twice the rotational frequency $\Omega$.
Such a signal is detectable only from sources in our galaxy or the
nearby ones (our Local Group). If the sensitivity of the detectors
is improved in the kHz region, then signals from the Virgo cluster
will be detectable. If the bar persists for many ($\sim$ 10-100)
rotation periods, then even signals from distances considerably
larger than the Virgo cluster will be detectable. The event rate
is of the same order as the SN rate (a few events per century per
galaxy): this means that given the appropriate sensitivity at
frequencies between 1-3kHz we might be able to observe a few
events per year. Bars can be also created during the merging of
NS-NS, BH-NS, BH-WD and even type II collapsars (see discussion in
\cite{Kobayasi2003}).

The above estimates in general, rely on Newtonian hydrodynamics
calculations; GR enhances the onset of the instability,
$\beta_{\rm dyn}\sim 0.24$ \cite{SBS2000} and $\beta_{\rm dyn}$
may be even lower for large values of the compactness (larger
$M/R$). The bar-mode instability may be excited for significantly
smaller $\beta$ if centrifugal forces produce a peak in the
density off the source's rotational center\cite{Centrella2001}.
Rotating stars with a high degree of differential rotation are
also dynamically unstable for significantly lower $\beta_{\rm
dyn}\gtrsim 0.01$ \cite{Shibata2003}. According to this scenario
the unstable neutron star settles down to a non-axisymmetric
quasi-stationary state which is a strong emitter of quasi-periodic
gravitational waves
\begin{equation}
h_{\rm eff} \approx 3\times 10^{-22} \left({R_{\rm eq} \over 30
{\rm km}} \right) \left({f\over 800 {\rm Hz}}\right)^{1/2}
\left({100 {\rm Mpc} \over d}\right) M_{1.4}^{1/2} .
\end{equation}
The bar-mode instability of differentially rotating neutron stars
is an excellent source of gravitational waves but it is based on
the assumption that the dissipation of non-axisymmetric
perturbations by viscosity and magnetic fields is negligible.
Magnetic fields might actually enforce the uniform rotation of the
star on a dynamical timescale and the persistent non-axisymmetric
structure might not have time to develop at all.

Numerical simulations have shown that the $m=1$ one-armed spiral
mode might become dynamically unstable for considerably lower
rotational rates \cite{Centrella2001, SBM2003}. The $m=1$
instability depends critically on the softness of the equation of
state (EoS) and the degree of differential rotation.

\subsection{CFS instability, f and r-modes}

After the initial bounce, neutron stars may maintain a
considerable amount of deformation. They settle down to an
axisymmetic configuration mainly due to emission of GWs, viscosity
and magnetic fields. During this phase QNMs are excited.
Technically speaking, an oscillating non-rotating star has equal
values $\pm |\sigma|$ (the frequency of a mode) for the forward
and backward propagating modes (corresponding to $m=\pm|m|$).
Rotation changes the mode frequency by an amount $\delta \sigma
\sim m \Omega$ and both the prograde and retrograde modes will be
dragged forward by the stellar rotation. If the star spins
sufficiently fast, this mode will appear moving forwards in the
inertial frame (an observer at infinity), but still backwards in
the rotating frame (an observer rotating with the star). Thus, an
inertial observer sees GWs with positive angular momentum emitted
by the retrograde mode, but since the perturbed fluid rotates
slower than it would in absence of the perturbation, the angular
momentum of the mode itself is negative. The emission of GWs
consequently makes the angular momentum of the mode increasingly
negative thus leading to an instability. From the above, one can
easily conclude that a mode will be unstable if it is retrograde
in the rotating frame and prograde for a distant observer
measuring a mode frequency $\sigma-m\Omega$ i.e. the criterion
will be $ \sigma(\sigma-m\Omega) < 0$.

This class of {\em frame-dragging instabilities} is usually
referred to as Chandrasekhar-Friedman-Schutz (CFS) instabilities.
For the high frequency ($f$ and $p$) modes this is possible only
for large values of $\Omega$ or for quite large $m$. In general,
for every mode there will always be a specific value of $m$ for
which the mode will become unstable, although only modes with
$|m|<5$ have an astrophysically significant growth time. The CFS
mechanism is not only active for fluid modes but also for the {\em
spacetime} or the so called {\em w}-modes\cite{KRA2002}. It is
easy to see that the CFS mechanism is not unique to gravitational
radiation: any radiative mechanism will have the same effect.

In GR, the {\em f-mode} ($l=m=2$) becomes unstable for $\beta
\approx 0.06-0.08$ \cite{SF1998}. If the star has significant
differential rotation the instability is excited for somewhat
higher values of $\beta$ (see e.g. \cite{Yoshida,
Stergioulas2003}). The $f$-mode instability is an excellent source
of GWs. After the brief dynamical phase, the PNS becomes unstable
and the instability deforms the star into a non-axisymmetric
configuration via the $l=2$ bar mode. Since the star loses angular
momentum, it spins-down, and the GW frequency sweeps from 1kHz
down to about 100Hz. Such a signal if properly modelled can be
detected from a distance of 100Mpc (if the mode grows to a large
nonlinear amplitude).

Rotation does not only shifts the spectra of these modes; it also
gives rise to a new type of restoring force, the Coriolis force,
with an associated new family of {\em rotational } or {\em
inertial} modes. Inertial modes are primarily velocity
perturbations and of special interest is the quadrupole inertial
mode ($r$-mode) with $l=m=2$. The frequency of the $r$-mode in the
rotating frame of reference is $\sigma = 2 \Omega /3$. Using the
CFS criterion for stability we can easily show that the $r$-mode
is unstable for any rotation rate of the star. For temperatures
between $10^{7}-10^{9}$K and rotation rates larger than 5-10\% of
the Kepler limit, the growth time of the unstable mode is smaller
than the damping times of the bulk and shear viscosity. The mode
grows until it saturates due to non-linear effects. The amplitude
of the emitted GWs depends on $\alpha$. Mode coupling might not
allow the growth of the instability to high amplitudes $\alpha
\approx 10^{-2}-10^{-3}$ \cite{Arras2002}. The existence of a
crust or of hyperons in the core \cite{LO2002}  and strong
magnetic fields, affect the efficiency of the instability (for an
extended review see \cite{Nils2003}). For newly-born neutron stars
the amplitude of GWs might not be large enough and the signals
will be detectable only from the local group of galaxies (
$d<1$Mpc)
\begin{equation}
h(t)\approx 10^{-21} \alpha \left( {\Omega \over 1 {\rm
kHz}}\right)\left({100 {\rm kpc} \over d}\right)
\end{equation}
If the compact object is a strange star, then the instability will
not reach high amplitudes ($\alpha \sim 10^{-3}-10^{-4}$) but it
will persist for a few hundred years and in this case there might
be up to ten unstable stars per galaxy at any time \cite{AJK2002}.
Integrating data for a few weeks can lead to an effective
amplitude $h_{\rm eff}\sim 10^{-21}$ for galactic signals at
frequencies $\sim 700-1000$Hz. The frequency of the signal changes
only slightly on a timescale of a few months, so the radiation is
practically monochromatic.

Old accreting neutron stars, radiating GWs due to the $r$-mode
instability, at frequencies 400-700Hz, are probably a better
source \cite{AKS1999, AJKS2000, Heyl, Wagoner}. Still, the
efficiency and the actual duration of the process depends on the
saturation amplitude $\alpha$. If the accreting compact object is
a strange star then it might be a persistent source which radiates
GWs for as long as accretion lasts\cite{AJK2002}.

\section{Oscillations of black holes and neutron stars}
%%%%%%%%%%%%%

{\em Black-hole ringing}. If the collapse produces a black hole
(collapsar type I or II) the black hole will ring until it settles
down to the stationary Kerr state. Although the ringing phase does
not last too long (a few tenths of msecs), the ringing due to the
excitation by the fallback material might last for secs. The
frequency and the damping time of the oscillations for the $l=m=2$
mode can be estimated via the relations \cite{Echeverria}
\begin{eqnarray}
\sigma&\approx& 3.2 {\rm kHz} \
M_{10}^{-1}\left[1-0.63(1-a/M)^{3/10}\right] \\
 Q&=&\pi \sigma
\tau \approx 2\left(1-a\right)^{-9/20} \label{bhqnm}
\end{eqnarray}
These relations together with similar ones either for the 2nd QNM
or the $l=2$, $m=0$ can uniquely determine the mass $M$ and
angular momentum $a$ of the BH if the frequency and the damping
time of the signal have been accurately extracted \cite{Finn,
Dryer}. The amplitude of the ring-down waves depends on the BH's
initial distortion. If the excitation of the BH is due to falling
material then the energy is roughly $\Delta E \gtrsim \epsilon \mu
c^2(\mu/M)$ where $\epsilon \gtrsim 0.01$ \cite{Davis1971}. This
leads to an effective GW amplitude
\begin{equation}
h_{\rm eff}\approx 2\times 10^{-21}\left({\epsilon \over 0.01}
\right)\left({10 {\rm Mpc}\over d}\right)\left( {\mu\over
M_\odot}\right)
\end{equation}

\medskip
{\em Neutron star ringing}. If the collapse leaves behind a
compact star, various types of oscillation modes might be excited
which can help us estimate parameters of the star such as radius,
mass, rotation rate and EoS\cite{AK1998, KAA2001, AC2001}. This
{\em gravitational wave asteroseismology} is a unique way to find
the radius and the EoS of compact stars. One can derive
approximate formulas in order to connect the observable
frequencies and damping times of the various stellar modes to the
stellar parameters. For example, for the fundamental oscillation
($l=2$) mode ($f$-mode) of non-rotating stars we get \cite{AK1998}
\begin{eqnarray}
\sigma({\rm kHz})&\approx& 0.8+1.6 M_{1.4}^{1/2}R_{10}^{-3/2}
+ \delta_1 m{\bar \Omega} \\
\tau^{-1}({\rm secs}^{-1})&\approx&
M_{1.4}^3R_{10}^{-4}\left(22.9-14.7M_{1.4}R_{10}^{-1}\right)+
\delta_2 m {\bar \Omega}
\end{eqnarray}
where ${\bar \Omega}$ is the normalized rotation frequency of the
star, and $\delta_1$ and $\delta_2$ are constants estimated by
sampling data from various EoS. The typical frequencies of the NS
modes are larger than 1kHz. On the other hand, 2D simulations of
rotating core-collapse have shown that if a rapidly rotating NS is
created, then the dominant mode is the quasi-radial mode
(``$l=0$''), radiating through its $l=2$ piece at frequencies
$\sim$  800Hz-1kHz \cite{DFM2002b}. Since each type of mode is
sensitive to the physical conditions where the amplitude of the
mode is greatest,  the more information we get from the various
types of modes the better we understand the details of the star.

Concluding, we should mention that the tidal disruption of a NS by
a BH \cite{Vallisneri} or the merging of two NSs \cite{Rasio} may
give valuable information for the radius and the EoS if we can
recover the signal at frequencies higher than 1 kHz.

This work has been supported by the EU Programme 'Improving the
Human Research Potential and the Socio-Economic Knowledge Base'
(Research Training Network Contract HPRN-CT-2000-00137).

\end{document}